\documentclass[english,aps,pra,reprint,nofootinbib,a4paper]{revtex4-1}
\usepackage[english]{babel}
\usepackage{amsfonts}
\usepackage{amsmath}
\usepackage{amssymb}
\usepackage{graphicx}
\usepackage{epstopdf}
\usepackage{enumerate}
\usepackage{color}

\begin{document}

\def\arXiv#1#2#3#4{{#1} #2 #3 {\it Preprint} #4}
\def\Book#1#2#3#4#5{{#1}, {\it #3} (#4, #5, #2).}
\def\Bookwd#1#2#3#4#5{{#1} {\it #3} (#4, #5, #2)}
\def\Journal#1#2#3#4#5#6#7{#1, #3, #4 \textbf{#5}, #6 (#2).}
\def\JournalE#1#2#3#4#5#6{#1, #3, #4 \textbf{#5}, #6 (#2).}
\def\eref#1{(\ref{#1})}

\newcommand{\dd}{\mbox{d}}
\newcommand{\EE}{\mathbb{E}}
\newcommand{\NN}{\mathbb{N}}
\newcommand{\PP}{\mathbb{P}}
\newcommand{\RR}{\mathbb{R}}
\newcommand{\TT}{\mathbb{T}}
\newcommand{\ZZ}{\mathbb{Z}}
\newcommand{\uu}{\mathbf{1}}
\newcommand{\HH}{\mathcal{H}}

\title{Quantum and random walks as universal generators of probability distributions}
\author{Miquel Montero}
\email{miquel.montero@ub.edu}
\affiliation{Secci\'o de F\'{\i}sica Estad\'{\i}stica i Interdisciplin\`aria, Departament de F\'{\i}sica de la Mat\`eria Condensada, Universitat de Barcelona (UB), Mart\'{\i} i Franqu\`es 1, E-08028 Barcelona, Spain}
\affiliation{Universitat de Barcelona Institute of Complex Systems (UBICS), Universitat de Barcelona, Barcelona, Spain}
\date{\today}


\begin{abstract}
Quantum walks and random walks bear similarities and divergences. One of the most remarkable disparities affects the probability of finding the particle at a given location: typically, almost a flat function in the first case and a bell-shaped one in the second case. Here I show how one can impose any desired stochastic behavior (compatible with the continuity equation for the probability function) on both systems by the appropriate choice of time- and site-dependent coins. This implies, in particular, that one can devise quantum walks that show diffusive spreading without loosing coherence, as well as random walks that exhibit the characteristic fast propagation of a quantum particle driven by a Hadamard coin. 

\end{abstract}
\maketitle

\section{Introduction}

Quantum walks~\cite{VA12} and random walks~\cite{GW94} have a long list of affinities and disparities. One can found the (now mostly deprecated) mixed expression ``quantum random walks'' in the first references exploring these new processes~\cite{ADZ93,TM02,NK03,JK03}, because they were developed as the quantum variants of the discrete random walk in one dimension: the Markov process in which, at every time step, a particle moves (either leftward or rightward) to one of the two neighboring sites as a result of the random outcome of a coin toss. The quantum particle however, like in the renowned case of the double-slit experiment, moves to both directions simultaneously, and this propagation takes place in a {\em deterministic\/} way: the wave function describing the system evolves unambiguously according to the value of some inner binary property |as, e.g., the spin or the chirality| whose state is locally updated by the action a unitary operator, known as the coin operator. Therefore, in this case, the location of the particle at a given instant of time is a probabilistic magnitude due to the intrinsic uncertainty inherent in every quantum phenomenon.

One of the first coin operators considered in the quantum-walk literature is the Hadamard coin~\cite{ABNVW01}, a real-valued unitary operator that performs a 
Hadamard transformation on the chirality of the particle. Since all the probabilities associated with this transformation are identical, the Hadamard walk can be considered as the quantum counterpart of a random walk with a fair coin. In both cases, the occupation probabilities of the more distant (although accessible) locations are exponentially small. But, while the central part of the distribution of the unbiased random walk quickly converges to a Gaussian, the location of a particle doing a Hadamard walk after $t$ steps is almost uniformly distributed in the range $[-t/\sqrt{2},t/\sqrt{2}]$, centered around the initial position of the particle, and therefore the quantum walker connects this point with any site within this interval after a lapse of time that is thus proportional to their relative distance. To perform the same operation, the unbiased random walker needs an amount of time that grows quadratically with the separation between the sites. 

These two diverging statistical traits are sometimes seen as paradigms of the two processes. The truth, however, is that these properties depend strongly on how the coin (operator) is chosen, and correspond to the homogeneous, time-independent (Markovian) case. Researchers have relaxed these conditions in the past and detected the emergence of new features in the system as, e.g., Anderson localization. Thus, among the publications on quantum walks, one can find examples of processes whose evolution is driven by site-dependent coins~\cite{BFT08,SK10,KW11,CGMV12,KLS13,ZXT14,XQTS14,AZ16}, time-dependent coins~\cite{RMM04,BNPRS06,AR09a,AR09b,MM14}, history-dependent coins~\cite{FAJ04,RBG13,SWH14}, and even {\em random\/} coins, unitary operators which are randomly chosen~\cite{JM10,AJ11,AVWW11,ACMSWW12,AJ12}. The lack of homogeneity is also a recurrent topic in the random-walk literature~\cite{RM01,FS07,LRT13,MPW17}.

My goal in this paper is, in a sense, just the opposite: starting from a given probability function, I want to deduce what is the proper coin selection to retrieve this distribution. With this aim, I consider here the discrete-time evolution of a particle moving on the integers as a result of the interaction with a set of site- and time-dependent (either quantum or random) coins. In a previous work~\cite{MM16}, I examined a particular instance of this problem, the design of a quantum walk that shown a binomial probability function, the distribution of a random walk with a fair coin. Here, I am going to generalize these results in {\em both directions\/}: I will find quantum walks with classical distributions, as well as random walks with quantum-like properties, provided that the comparison is limited to their common probabilistic aspects. 

The paper is organized as follows. Section~\ref{Sec_QW} considers the case of the discrete-time quantum walk on the line with a time- and site-dependent coin operator. Section~\ref{Sec_RW} is devoted to random walks with the same degree of freedom in its design. In  Sec.~\ref{Sec_Uniform} I show how one can recover a uniform probability function in both cases. Section~\ref{Sec_Inter} explores the possibility of interchanging the traditional roles of the two processes. The paper ends with Sec.~\ref{Sec_Conclusion}, where conclusions are drawn, while some technical discussion is left for the appendices.

\section{QW with desired distribution}
\label{Sec_QW}

Let us begin with the fundamentals of the quantum-mechanical side of the problem. As I announced previously, along this paper we will identify particle positions through integer numbers, so let us call $\HH^{\text p}$ the associated Hilbert space, with the usual span $\left\{| n\rangle : n \in \ZZ\right\}$. $\HH^{\text c}$ will represent the Hilbert space of coin states and $\left\{|+\rangle, |-\rangle\right\}$ its orthogonal basis. The mathematical representation of the state of our discrete-time, discrete-space quantum walk resides in the tensor-product space $\HH\equiv\HH^{\text c}\otimes \HH^{\text p}$ and changes as a result of the action of the evolution operator $\widehat{T}_t$ on it: $\widehat{T}_t\equiv \widehat{S}\, \widehat{U}_t$, where the {\em coin\/} $\widehat{U}_t$ is a time- and site-dependent, real-valued unitary operator of the form:
\begin{eqnarray}
\widehat{U}_t
&\equiv& \sum_{n=-\infty}^{\infty}\big[\cos \theta^{ }_{n,t} |+\rangle  \langle +| + \sin \theta^{ }_{n,t} |+\rangle  \langle -| \nonumber \\
&+&\sin \theta^{ }_{n,t}  |-\rangle  \langle +| - \cos \theta^{ }_{n,t}  |-\rangle  \langle -|\big]\otimes |n\rangle  \langle n|,
\label{U_coin_gen}
\end{eqnarray}
with $0\leq \theta^{ }_{n,t}\leq \pi$, and $\widehat{S}$ is the operator that {\em shifts\/} the walker position according to the coin component of the state vector:
\begin{equation}
\widehat{S} |\pm\rangle\otimes| n\rangle = |\pm\rangle\otimes| {n\pm 1}\rangle.
\end{equation}
In the discrete-time version of quantum (and random) walks, time increases in regular ticks, so one can adjust time units so that $t$ becomes an integer variable: the state of the system at a later time, $|\psi\rangle_{t+1}$, is recovered after applying $\widehat{T}_t$ to $|\psi\rangle_{t}$:   
\begin{equation}
|\psi\rangle_{t+1} =\widehat{T}_t|\psi\rangle_{t}.
\label{evol_t}
\end{equation}
Equation~\eqref{evol_t} leads to the following set of recursive equations that fully characterizes the dynamics of the system,
\begin{eqnarray}
\psi^{ }_{+}(n+1,t+1)&=&\cos \theta^{ }_{n,t} \,\psi^{ }_{+}(n,t)+\sin \theta^{ }_{n,t} \,\psi^{ }_{-}(n,t),\nonumber\\\label{Rec_P}
\\
\psi^{ }_{-}(n-1,t+1)&=&\sin \theta^{ }_{n,t} \,\psi^{ }_{+}(n,t)-\cos \theta^{ }_{n,t} \,\psi^{ }_{-}(n,t),\nonumber\\
\label{Rec_M}
\end{eqnarray}
expressed in terms of the wave-function components, $\psi_{\pm}(n,t)$, the projections of the state of the walker into the elements of the basis of the Hilbert space:
\begin{eqnarray}
\psi^{ }_{+}(n,t)&\equiv& \langle n|  \otimes  \langle +| \psi\rangle_t, \label{Def_Psi_P}\\
\psi^{ }_{-}(n,t)&\equiv& \langle n|  \otimes  \langle -| \psi\rangle_t. \label{Def_Psi_M} 
\end{eqnarray}
We will assume that the particle is initially located at the origin, $\psi^{ }_{\pm}(n,0)=0$ if $n\neq0$, implying this that $\psi^{ }_{\pm}(n,t)=0$ for $\left|n\right|>t$, in general. We also assume that the wave function is real. 
The reason behind considering real-valued magnitudes is to clearly ensure that quantum walks and random walks to be introduced here share the same number of degrees of freedom. The viability of approaches to this same issue based on complex-valued operators and wave functions are not discarded, however.

My first aim is to show how a quantum experiment can be designed with custom probabilistic properties |as long as the null sets are kept unchanged. So, let us introduce $\rho(n,t)$, the likelihood of finding the particle in a particular position $n$ at a given time $t$, the probability function. In the case of a quantum walker, this probability is recovered through the wave-function components:
\begin{equation}
\rho(n,t)\equiv\psi^{2}_{+}(n,t)+\psi^{2}_{-}(n,t).
\label{Constraint_old}
\end{equation}
The free parameters that determine the features of the coin operators are in this case the angular variables $\theta^{ }_{n,t}$. Therefore, one has as many unknown quantities as independent equations,~\footnote{After a raw inspection, it could be concluded that in this problem the number of unknown quantities exceeds the number of constraints and that the system of equations is underdetermined: after all, different choices for $\psi^{ }_{+}(n,t)$ and $\psi^{ }_{-}(n,t)$ may be (in principle) congruent with the same value of $\rho(n,t)$. This is not true here, with only one marginal exception: $\psi^{ }_{+}(0,0)$ and $\psi^{ }_{-}(0,0)$ are arbitrary, provided that $\rho(0,0)=1$. The adequate choice for $\theta^{ }_{0,0}$ is recovered from Eqs.~\eqref{Cos_Gen} and~\eqref{Sin_Gen} below.} so, it is not surprising that our objective can be readily attained. To this end, let us begin by focusing our attention on the conditions that the wave-function components must satisfy for ensuring the self-consistency of the problem. From Eqs.~\eqref{Rec_P} and~\eqref{Rec_M} one gets
\begin{equation}
\psi^{2}_{+}(n+1,t+1)+\psi^{2}_{-}(n-1,t+1)=\rho(n,t),
\label{Constraint_new}
\end{equation}
but, at the same time, cf. Eq.~\eqref{Constraint_old},
\begin{equation}
\psi^{2}_{+}(n+1,t+1)+\psi^{2}_{-}(n+1,t+1)=\rho(n+1,t+1).
\label{Constraint_old_revisited}
\end{equation}
In particular, for $n=t$, $t\geq 1$, one has $\psi^{ }_{-}(t,t)=0$, see Eq.~\eqref{Rec_M}, and therefore 
\begin{equation}
\psi^{2}_{+}(t+1,t+1)=\rho(t+1,t+1).
\end{equation}
This means that by subtracting Eq.~\eqref{Constraint_old_revisited} from Eq.~\eqref{Constraint_new} for $n=t$ one gets
\begin{equation}
\psi^{2}_{-}(t-1,t+1)=\rho(t,t)-\rho(t+1,t+1).
\end{equation}
This result can be used to compute $\psi^{2}_{+}(t-1,t+1)$ through Eq.~\eqref{Constraint_old_revisited}, and we may continue with this reasoning until obtaining the general rule, valid for $t\geq1$,
\begin{eqnarray}
\psi^{2}_{+}(n,t)&=&\sum_{m=n}^{t}\rho(m,t)-\sum_{m=n+1}^{t-1}\rho(m,t-1),\label{Psi_P_Gen}\\
\psi^{2}_{-}(n,t)&=&\sum_{m=n+1}^{t-1}\rho(m,t-1)-\sum_{m=n+2}^{t}\rho(m,t).\label{Psi_M_Gen}
\end{eqnarray}
Since $\psi^{2}_{+}(n,t)$ and $\psi^{2}_{-}(n,t)$ are positive-definite magnitudes, in practice, this introduces a constraint on the \emph{evolution} of $\rho(n,t)$. This limitation does not stem from our assumption that the wave function is real valued, however: It has its origin in the nearest-neighbor restriction of both quantum and random walker dynamics |see App.~\ref{App_A} for a more detailed discussion.

Satisfied this requirement, it can be checked how Eqs.~\eqref{Constraint_old} and~\eqref{Constraint_new} are fulfilled, as well as the boundary conditions: $\psi^{ }_{+}(-t,t)=0$, and $\psi^{2}_{-}(-t,t)=\rho(-t,t)$. Alternatively, one can show the soundness of the solution by induction. 
Now we can use either Eq.~\eqref{Rec_P} or Eq.~\eqref{Rec_M} to finally find
\begin{eqnarray}
\cos \theta^{ }_{n,t} &=& \frac{\psi^{ }_{+}(n,t) \psi^{ }_{+}(n+1,t+1)}{\rho(n,t)}\nonumber\\
&-& \frac{\psi^{ }_{-}(n,t) \psi^{ }_{-}(n-1,t+1)}{\rho(n,t)},\label{Cos_Gen}
\end{eqnarray}
\begin{eqnarray}
\sin \theta^{ }_{n,t} &=& \frac{\psi^{ }_{-}(n,t) \psi^{ }_{+}(n+1,t+1)}{\rho(n,t)}\nonumber\\
&+& \frac{\psi^{ }_{+}(n,t) \psi^{ }_{-}(n-1,t+1)}{\rho(n,t)}\label{Sin_Gen}.
\end{eqnarray}

Equation~\eqref{Constraint_new} reflects the law of probability conservation. Rearranging this expression, one can see how the same statement can be also expressed as follows
\begin{eqnarray}
\rho(n,t)&=&\frac{1}{2}\left[\rho(n-1,t-1)+J(n-1,t-1)\right.\nonumber\\
&+&\left.\rho(n+1,t-1)-J(n+1,t-1)\right],
\label{Rec_rho_QW}
\end{eqnarray}
where $J(n,t)$ is the net flux of probability leaving site $n$
\begin{eqnarray}
J(n,t)&\equiv& \psi^{2}_{+}(n+1,t+1)-\psi^{2}_{-}(n-1,t+1)\nonumber \\
&=& \cos 2\theta^{ }_{n,t}\left[\psi^{2}_{+}(n,t)-\psi^{2}_{-}(n,t)\right]\nonumber\\
&+&2\sin 2\theta^{ }_{n,t}\psi^{ }_{+}(n,t) \psi^{ }_{-}(n,t),
\label{J_QW}
\end{eqnarray}
a {\em vectorial\/} quantity: it is positive if there is a net flux of probability to larger values of $n$, and negative otherwise.~\footnote{Since we have a bipartite graph, all the probability leaves the site after every clock tick: i.e., if $\rho(n,t) \neq 0$, then $\rho(n,t+1) = 0$. $J(n,t)$ is not the difference of these two quantities.} This magnitude is very useful in subsequent derivations, as we will see below.  

\section{RW with desired distribution}
\label{Sec_RW}

The inhomogeneous, time-dependent random walk, $X_t$, is a non-Markovian process whose one-step evolution can be expressed as follows: If at time $t$ the walker is at a given location, $X_{t}=n$, then at time $t+1$ one has
\begin{equation}
X_{t+1}=\left\{
\begin{array}{ll}
n+1,&\mbox{ with probability } p_{n,t},\\
n-1,&\mbox{ with probability } (1-p_{n,t}).
\end{array}
\right.
\label{process}
\end{equation}
The corresponding recursive equation for the probability function reads:
\begin{eqnarray}
\rho(n,t)&=&\cos^2 \bar{\theta}^{ }_{n-1,t-1} \,\rho(n-1,t-1)\nonumber\\
&+&\sin^2 \bar{\theta}^{ }_{n+1,t-1} \,\rho(n+1,t-1),
\label{Rec_rho_RW}
\end{eqnarray}
where we have expressed $p_{n,t}$ as $p_{n,t}=\cos^2 \bar{\theta}^{ }_{n,t}$ for comparison purposes. From Eq.~\eqref{Rec_rho_RW} one can easily conclude the validity of expression~\eqref{Rec_rho_QW} also in this case, since now
\begin{equation}
J(n,t)\equiv \left(2p_{n,t}-1\right)\rho(n,t)=\cos 2\bar{\theta}^{ }_{n,t}\,\rho(n,t).
\label{J_RW}
\end{equation}

The general solution of the classical problem for arbitrary $\rho(n,t)$ can be attained, in this case, with the help of the $z$ transform,
\begin{eqnarray*}
\widehat{\rho}(z,t)&\equiv&\mathcal{Z}\left[\rho(n,t),n,z\right]=\sum_{n=-\infty}^{\infty} \rho(n,t)z^{-n},\\ 
\widehat{J}(z,t)&\equiv&\mathcal{Z}\left[J(n,t),n,z\right]= \sum_{n=-\infty}^{\infty} J(n,t)z^{-n}.
\end{eqnarray*}
Equation~\eqref{Rec_rho_QW} leads to
\begin{equation}
\widehat{J}(z,t)=\frac{2 z\widehat{\rho}(z,t+1)-(1+z^2)\widehat{\rho}(z,t)}{1-z^2},
\label{hat_J_sol}
\end{equation}
and therefore
\begin{equation}
\cos 2\bar{\theta} ^{ }_{n,t}=\frac{1}{\rho(n,t)}\mathcal{Z}^{-1}\left[\widehat{J}(z,t),z,n\right].
\label{cos_RW}
\end{equation}

\section{Uniform distribution}
\label{Sec_Uniform}

I will illustrate these ideas through a simple but paradigmatic example where closed expressions can be found. Consider, for instance, the uniform distribution:
\begin{equation}
\rho(n,t)=\frac{1}{t+1},
\label{rho_uniform}
\end{equation}
for $n\in\left\{-t,-t+2,\cdots,t-2,t\right\}$. Equations~\eqref{Psi_P_Gen} and ~\eqref{Psi_M_Gen} lead to
\begin{eqnarray}
\psi^{ }_{+}(n,t)&=& \sqrt{\frac{t+n}{2 t (t+1)}}, \label{Psi_P_Flat}\\
\psi^{ }_{-}(n,t)&=&  \sqrt{\frac{t-n}{2 t (t+1)}}, \label{Psi_M_Flat} 
\end{eqnarray}
and correspondingly
\begin{eqnarray}
\cos \theta^{ }_{n,t} &=&\frac{1}{2}\sqrt{\frac{(t+n)(t+n+2)}{t(t+2)}} \nonumber\\
&-& \frac{1}{2}\sqrt{\frac{(t-n)(t-n+2)}{t(t+2)}},\label{Cos_Flat}\\
\sin \theta^{ }_{n,t} &=&\frac{1}{2}\sqrt{\frac{(t-n)(t+n+2)}{t(t+2)}} \nonumber\\
&+& \frac{1}{2}\sqrt{\frac{(t+n)(t-n+2)}{t(t+2)}}.\label{Sin_Flat}
\end{eqnarray}

We can use the results above to assess the value of $J(n,t)$, 
\begin{equation}
J(n,t)=\frac{n}{(t+1)(t+2)}.
\label{J_uniform}
\end{equation}
Provided with this information, we can solve the classical problem without passing through Eq.~\eqref{hat_J_sol} in this case:~\footnote{
The explicit functional forms of $\widehat{\rho}(z,t)$ and $\widehat{J}(z,t)$ for this case are:
\begin{eqnarray*}
\widehat{\rho}(z,t)&=&\frac{z^{-t}}{t+1}\frac{1-z^{2 t+2}}{1-z^2},\\
\widehat{J}(z,t)&=&\frac{z^{-t}}{(t+1)(t+2)}\frac{t(1+z^2)(1+z^{2 t+2})-2z^2(1-z^{2t})}{(1-z^2)^2}.
\end{eqnarray*}
The surprisingly disparity in the complexity of these formulas when compared to Eqs.~\eqref{rho_uniform} and~\eqref{J_uniform} is in great measure due to the fact that the last expressions only apply for alternating sites, i.e., $n\in\left\{-t,-t+2,\cdots,t-2,t\right\}$, being zero otherwise.
} 
recall that $J(n,t)$ is the same in both flavors of the walk, so we can substitute~\eqref{rho_uniform} and~\eqref{J_uniform} in Eq.~\eqref{cos_RW} to find
\begin{equation}
\cos 2\bar{\theta}^{ }_{n,t}=\frac{J(n,t)}{\rho(n,t)}=\frac{n}{t+2},
\end{equation}
that implies
\begin{eqnarray}
p_{n,t}&=&\cos^2\bar{\theta} ^{ }_{n,t}=\frac{1}{2}\left(1+\frac{n}{t+2}\right).
\end{eqnarray}

\section{Interchanging roles}
\label{Sec_Inter}

Finally, I want to explore the possibility of role interchange. In a recent work~\cite{MM16}, I considered the case in which the $\rho(n,t)$ corresponding to a time- and site-dependent quantum walk matched the probability function of a site-homogeneous Markovian random walk, i.e., the binomial distribution:
\begin{eqnarray}
\rho(n,t)&=&\frac{t!}{\left(\frac{t+n}{2}\right)! \left(\frac{t-n}{2}\right)!}p^{\frac{t+n}{2}}\left(1-p\right)^{\frac{t-n}{2}},
\label{Rho_RW}
\end{eqnarray}
for $n\in\left\{-t,-t+2,\cdots,t-2,t\right\}$. There it was shown that the solution for this problem reads 
\begin{eqnarray}
\psi^{ }_{+}(n,t)&=& \sqrt{p} \sqrt{\rho(n-1,t-1)}, \label{Redund_Psi_P}\\
\psi^{ }_{-}(n,t)&=&  \sqrt{1-p} \sqrt{ \rho(n+1,t-1)}, \label{Redund_Psi_M} 
\end{eqnarray}
two expressions whose suitability can be checked by direct insertion in Eqs.~\eqref{Constraint_old} and~\eqref{Constraint_new}. Alternatively, it is very elucidative the computation of $J(n,t)$, since in this case
\begin{eqnarray}
J(n,t)=(2p-1)\rho(n,t),
\label{J_CQW}
\end{eqnarray} 
what corresponds to the flux of probability of a random walk with a constant jump likelihood, cf. Eq.~\eqref{J_RW}.

Here, I will examine the opposite situation: how a time- and site-dependent random walk can mimic the characteristic properties of a {\em standard\/} quantum walk. In particular, we are going to focus our attention on the celebrated Hadamard walk, for which $\theta^{ }_{n,t}=\pi/4$. This means that, on the one side, see Eq.~\eqref{J_QW},
\begin{equation}
J(n,t)= 2\psi^{ }_{+}(n,t) \psi^{ }_{-}(n,t),
\end{equation}
and, on the other side, see Eq.~\eqref{J_RW},
\begin{eqnarray}
p_{n,t}=\frac{\rho(n,t)+J(n,t)}{2 \rho(n,t)},
\end{eqnarray}
that is
\begin{equation}
p_{n,t}=\frac{\left[\psi^{ }_{+}(n,t)+\psi^{ }_{-}(n,t)\right]^2}{2\rho(n,t)}.
\label{Hadamard_RW}
\end{equation}

\begin{figure}[htbp]
\includegraphics[width=0.9\columnwidth,keepaspectratio=true]{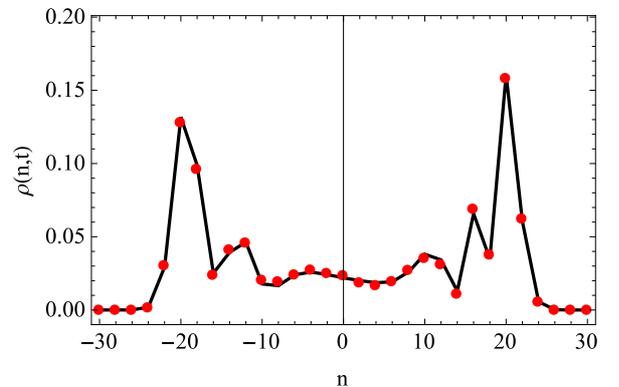}\
\caption{Probability functions at $t=30$. The red dots were obtained by averaging 10\,000 simulated trajectories of an inhomogeneous, time-dependent random walk. The solid black line corresponds to the probability function of a quantum walk with a Hadamard coin. Only even locations are shown as the probability is zero for odd sites.} 
\label{Fig_Hadamard}
\end{figure}

Closed expressions for the wave-function components of plain quantum walks (including Hadamard walks) are unwieldy but available |see, e.g., App.~\ref{App_B} or Ref.~\cite{MM15}. In Fig.~\ref{Fig_Hadamard} we can observe the almost perfect correspondence between the probability function of the random walk with inhomogeneous probabilities, and the one of the Hadamard walk with initial state:
\begin{equation}
|\psi\rangle_{0}=\left[\frac{\sqrt{2-\sqrt{2}}}{2}|+\rangle + \frac{\sqrt{2+\sqrt{2}}}{2}|-\rangle\right] \otimes| 0\rangle.
\end{equation}
This apparently capricious choice was made to get a quasi-symmetrical $\rho(n,t)$ \cite{NK05,BP07,MM15,BSJ15,SJ16}. Full symmetry in quantum walks endowed with a real-valued, homogeneous coin operator, as in the case of a Hadamard walk, necessarily involves the use of complex coefficients for describing the initial coin state \cite{JK03,MM15}.~\footnote{This does not represent a restriction: we can use this method to replace a quantum walk with a homogeneous coin operator in the complex plane by either a random or quantum walk on the reals with an inhomogeneous coin |see App.~\ref{App_B}.} 

\section{Conclusion}
\label{Sec_Conclusion}

In this paper, I have shown how a time- and site-dependent coin is an extremely useful and versatile tool for the design of both quantum and random walks on the line. Such approach entails enough generality to give rise to any desired probabilistic fingerprint either through quantum or classical randomness: I have deduced the rules that must be employed for unambiguously  assessing the values of the parameters that fully determine the evolution of the two kind of systems.

This means, in particular, that the extra degree of freedom of the quantum walker associated with its chirality does not introduce further arbitrariness into the problem. This fact is not the consequence of the restriction that I have considered along the text by demanding that the Hilbert space of the quantum particle is defined on the reals rather than on the complex plane: Since a quantum walk with a time- and site-dependent coin operator taking values on the reals can mimic any desired probability function, it is also capable of reproducing the probabilistic behavior of general, complex-valued quantum walks. 

As a final remark, note that heterogeneity is a degree of freedom that could be eventually superfluous: This opens the possibility of finding new optimal simulating strategies, a task which is  left for a future research.

\acknowledgments
The author acknowledges support from the Spanish Ministerio de Econom\'{\i}a y Competitividad (MINECO) under Contract No. FIS2013-47532-C3-2-P, from the Spanish Agencia Estatal de Investigaci\'on and from the European Fondo Europeo de Desarrollo Regional  (AEI/FEDER, UE) under Contract No. FIS2016-78904-C3-2-P, and from the  Catalan Ag\`encia de Gesti\'o d'Ajuts Universitaris i de Recerca (AGAUR), Contract No. 2014SGR608.

\appendix

\section{Valid sequences of distributions}
\label{App_A}
A property shared by random and quantum walks on a line is that the particle can jump to one (or both) of the nearest sites. This means that the present probability transition matrix can only connect adjacent points. This condition is summarized in the probability conservation equation:
\begin{eqnarray}
\rho(n,t)&=&\frac{1}{2}\left[\rho(n-1,t-1)+J(n-1,t-1)\right.\nonumber\\
&+&\left.\rho(n+1,t-1)-J(n+1,t-1)\right],
\label{Rec_rho}
\end{eqnarray}
which proves valid in general: The net flux of probability leaving site $n$ at time $t$, $J(n,t)$, reads 
\begin{equation}
J(n,t)\equiv \left|\psi^{ }_{+}(n+1,t+1)\right|^2-\left|\psi^{ }_{-}(n-1,t+1)\right|^2,
\label{J_QW_App}
\end{equation}
in the complex-valued quantum case, and
\begin{equation}
J(n,t)\equiv \left(2p_{n,t}-1\right)\rho(n,t).
\label{J_RW_App}
\end{equation}
in the classical case. In both instances, one has the natural constraint
\begin{equation}
-\rho(n,t)\leq J(n,t)\leq \rho(n,t).
\label{J_bounds}
\end{equation}
Note that Eq.~\eqref{Rec_rho} can be rewritten as
\begin{eqnarray}
J(n+2,t) &=&J(n,t)+\rho(n,t)+\rho(n+2,t)\nonumber\\
&-&2\rho(n+1,t+1),
\label{Rec_J}
\end{eqnarray}
with the following boundary condition, if one assumes that the walker starts at the origin:
\begin{equation}
J(-t,t)= \rho(-t,t)-2\rho(-t-1,t+1),
\end{equation}
and where the usual restrictions apply, $\rho(n,t)=0$ for $|n|>t$. Therefore, given a certain sequence of probability functions $\rho(n,t)$, $t\geq 0$, it will correspond to the evolution of some physical system governed by our dynamics if and only if Eq.~\eqref{J_bounds} holds for every $J(n,t)$ thus obtained.

\section{Simulating QWs on the complex plane}
\label{App_B}

Let us consider the most general homogeneous QW on the line, driven by the unitary coin operator
\begin{eqnarray}
\widehat{U}
&\equiv& e^{i \chi}\Big[ e^{i \alpha} \cos \theta |+\rangle  \langle +| +e^{-i \beta} \sin \theta |+\rangle  \langle -| \nonumber \\
&+& e^{i \beta} \sin \theta  |-\rangle  \langle +| - e^{-i \alpha} \cos \theta  |-\rangle  \langle -|\Big]\otimes |n\rangle  \langle n|. \nonumber \\
\label{U_coin_gen_App}
\end{eqnarray}
The evolution operator $\widehat{T}\equiv \widehat{S}\, \widehat{U}$ induces in this case the following set of recursive equations in the wave function:
\begin{eqnarray}
\psi^{ }_{+}(n,t)&=&e^{i\chi^{ }_{}} \big[e^{i\alpha}\cos \theta \,\psi_{+}(n-1,t-1)\nonumber\\&+&
e^{-i\beta}\sin \theta \,\psi_{-}(n-1,t-1)\big],
\label{Rec_P_bas}
\end{eqnarray}
and
\begin{eqnarray}
\psi^{ }_{-}(n,t)&=&e^{i\chi} \big[e^{i\beta}\sin \theta \,\psi_{+}(n+1,t-1)\nonumber\\&-&
e^{-i\alpha}\cos \theta\,\psi_{-}(n+1,t-1)\big],
\label{Rec_M_bas}
\end{eqnarray}
whose general solution~\cite{MM15} can be written in a compact way by using $\psi_{+}(0,0)$ and $\psi_{-}(0,0)$,
\begin{eqnarray*}
\psi_{+}(0,0)&=&\cos \eta,\\
\psi_{-}(0,0)&=&e^{i\gamma}\sin \eta,
\end{eqnarray*}
and the non-zero components of the wave function at time $t=1$,  
\begin{eqnarray*}
\psi_{+}(+1,1)&=&e^{i \chi }\left[e^{i\alpha}\cos\eta\cos \theta+ e^{i(\gamma-\beta)}\sin \eta\sin \theta\right],\\
\psi_{-}(-1,1)&=&e^{i \chi }\left[e^{i \beta}\cos\eta\sin \theta-  e^{i(\gamma-\alpha)}\sin \eta\cos \theta\right],
\end{eqnarray*}
since on has $\psi_+(-1,1)=\psi_-(+1,1)=0$, cf. Eqs.~\eqref{Rec_P_bas} and~\eqref{Rec_M_bas}. In terms of the preceding quantities, and for $n\in\{-t,-t+2,\cdots,t-2,t\}$, the chiral components of the wave function read
\begin{eqnarray}
\psi_{+}(n,t)&=&e^{i(\chi\cdot t +\alpha\cdot n)}\Big[\psi_{+}(0,0)  \Lambda(n,t)\nonumber \\
&+&e^{-i(\chi +\alpha)}\psi_{+}(+1,1) \Lambda(n-1,t+1)\Big],
\label{Sol_Psi_P}
\end{eqnarray}
and
\begin{eqnarray}
\psi_{-}(n,t)&=&e^{i(\chi \cdot t -\alpha \cdot n)}\Big[\psi_{-}(0,0) \Lambda(n,t)\nonumber\\
&+& e^{-i(\chi -\alpha)} \psi_{-}(-1,1) \Lambda(n+1,t+1)\Big],
\label{Sol_Psi_M}
\end{eqnarray}
where
\begin{eqnarray}
\Lambda(n,t)&\equiv&\frac{1}{t+1}\Bigg\{\frac{1+(-1)^t}{2}\nonumber \\&+&
\sum_{r=1}^{t}  \frac{1}{\cos\omega_{r,t}}\cos\left[(t-1)\cdot \omega_{r,t}-\frac{\pi r n}{t+1}\right]\Bigg\},\nonumber \\
\label{Lambda_def}
\end{eqnarray}
and 
\begin{equation}
\omega_{r,t} \equiv \arcsin\left(\cos\theta \sin  \frac{\pi r}{t+1}\right).
\label{omega_r}
\end{equation}

Observe how  $\Lambda(n,t)$ does not depend on $\chi$, $\alpha$, $\beta$, $\gamma$ or $\eta$, it is a function of $\theta$ through the value of $\cos\theta$. The probability function  does not depend either on $\chi$; and $\alpha$, $\beta$ and $\gamma$ will appear in $\rho(n,t)$ only through the following combination $\varphi=\alpha+\beta-\gamma$. (Therefore, a common simplification made in the literature is considering that $\chi=\alpha=\beta=0$.) The asymptotic behavior of the probability function, for $|n|\lesssim t\cos\theta$, is known to be 
\begin{eqnarray}
\rho(n,t)&\to&\frac{2}{\pi} \frac{t}{t^2-n^2}\frac{\sin\theta}{\sqrt{t^2\cos^2\theta-n^2}} \nonumber \\
&\times&\Big[t+n \left(\cos 2\eta +\sin 2\eta\tan\theta \cos \varphi  \right)\Big],
\label{Main_Prob_Asymp}
\end{eqnarray}
a symmetric function around the origin whenever
\begin{eqnarray}
\cos 2\eta \cos \theta+\sin 2 \eta\sin \theta\cos\varphi=0.
\label{symm_A}
\end{eqnarray}
The selection used in the main text, $\theta=\pi/4$, $\eta=3\pi/8$, and $\varphi=0$ is one of the multiple solutions of this equation. Moreover, given  $\left\{\theta_0,\eta_0,\varphi_0\right\}$ such that
\begin{eqnarray}
\cos 2\eta_0 \cos \theta_0+\sin 2 \eta_0\sin \theta_0\cos\varphi_0=a_0,
\end{eqnarray}
with $|a_0|\leq1$, there is always an alternative choice $\left\{\theta_1=\theta_0,\eta_1,\varphi_1\right\}$ with $|\cos \varphi_1|=1$, leading to the same value of $a_0$.

Equation~\eqref{symm_A} is a necessary but not a sufficient condition in order to have exact symmetry in $\rho(n,t)$. In addition, one must demand that
\begin{equation}
\cos 2\eta \cos 2\theta+\sin 2 \eta\sin 2\theta\cos\varphi=0.
\label{symm_B}
\end{equation}
The richer solution of this set of equations, the one which does not requires that either $\cos \theta=0$ or $\sin \theta=0$, corresponds to  $\eta=\pi/4$ and $\varphi=\pm\pi/2$. In particular, for $\eta=\pi/4$ and $\gamma=\pi/2$ one has 
\begin{eqnarray*}
\psi_{+}(0,0)&=&\frac{1}{\sqrt{2}},\\
\psi_{-}(0,0)&=&\frac{i}{\sqrt{2}},\\
\psi_{+}(+1,1)&=&\frac{1}{\sqrt{2}}e^{i \theta},\\
\psi_{-}(-1,1)&=&\frac{-i}{\sqrt{2}}e^{i \theta},
\end{eqnarray*}
and therefore
\begin{eqnarray}
\psi_{+}(n,t)&=&\frac{1}{\sqrt{2}}\Big[ \Lambda(n,t)
+e^{i\theta}\Lambda(n-1,t+1)\Big],\\
\label{Sol_Psi_P_HW}
\psi_{-}(n,t)&=&\frac{i}{\sqrt{2}}\Big[\ \Lambda(n,t)
-e^{i\theta}  \Lambda(n+1,t+1)\Big].
\label{Sol_Psi_M_HW}
\end{eqnarray}
Since here
\begin{equation*}
\rho(n,t)=\left|\psi^{ }_{+}(n,t)\right|^2+\left|\psi^{ }_{-}(n,t)\right|^2,
\end{equation*}
one has
\begin{eqnarray}
\rho(n,t)&=&\frac{1}{2}\Lambda^2(n+1,t+1)+\frac{1}{2}\Lambda^2(n-1,t+1)\nonumber\\
&+&\Lambda(n,t)\Lambda(n,t+2),
\label{Rho_HW}
\end{eqnarray}
where the following general recursive formula for $\Lambda(n,t)$ has been used~\cite{MM15}:
\begin{eqnarray}
\Lambda(n,t)&=&\cos \theta \left[\Lambda(n+1,t+1)-\Lambda(n-1,t+1)\right]\nonumber\\
&+&\Lambda(n,t+2).
\label{Lambda_recursive}
\end{eqnarray}

\begin{figure}[htbp]
\includegraphics[width=0.9\columnwidth,keepaspectratio=true]{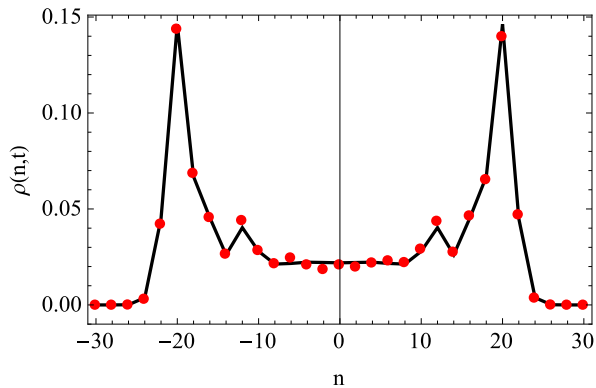}\
\caption{Probability functions at $t=30$. The red dots were obtained by averaging 10\,000 simulated trajectories of an inhomogeneous, time-dependent random walk. The solid black line corresponds to the probability function of a symmetric Hadamard walk. Only even locations are shown as the probability is zero for odd sites.} 
\label{Fig_Hadamard_App}
\end{figure}

Let us consider in the first place the inhomogeneous RW that mimics this stochastic evolution. The time- and site-dependent probability must be set in such a way that
\begin{equation*}
p_{n,t}=\frac{\rho(n,t)+J(n,t)}{2 \rho(n,t)}=\frac{\left|\psi^{}_{+}(n+1,t+1)\right|^2}{\rho(n,t)},
\end{equation*}
where
\begin{eqnarray*}
\left|\psi^{}_{+}(n+1,t+1)\right|^2&=&\frac{1}{2} \Lambda^2(n+1,t+1)+\frac{1}{2}\Lambda^2(n,t+2)\\
&+&\cos\theta \Lambda(n+1,t+1)\Lambda(n,t+2).
\end{eqnarray*}
In Fig.~\ref{Fig_Hadamard_App} we can see the good agreement between the analytic expression and the numerical results obtained after the simulation of 10\,000 trajectories, for $\theta=\pi/4$.

Finally, let us consider the design of an inhomogeneous QW on the reals with the same statistical properties of this symmetric Hadamard walk. In the remaining of this section we will denote by $\widetilde{\psi}^{}_{\pm}(n,t)$ the real-valued wave functions of this inhomogeneous QW, and by $\widetilde\theta^{ }_{n,t}$ the parameter associated to its coin operator, and we keep $\psi^{}_{\pm}(n,t)$ and $\theta$ for the homogeneous QW on the complex plane.

The solution of this problem is almost strightforward. We must simply choose the time- and site-dependent coin opearator for which 
\begin{eqnarray*}
\widetilde\psi^{}_{+}(n,t)&=&\sqrt{\left|\psi^{}_{+}(n,t)\right|^2},\\ 
\widetilde\psi^{}_{-}(n,t)&=&\sqrt{\left|\psi^{}_{-}(n,t)\right|^2},
\end{eqnarray*}
since in this case one has automatically granted the same value of $\rho(n,t)$.  But this demand can be readily accomplished |see Eqs.~(15) and~(16) in the main text| through
\begin{eqnarray*}
\cos \widetilde\theta^{ }_{n,t} &=& \frac{ \widetilde\psi^{ }_{+}(n,t)  \widetilde\psi^{ }_{+}(n+1,t+1)}{\rho(n,t)}\nonumber\\
&-& \frac{ \widetilde\psi^{ }_{-}(n,t)  \widetilde\psi^{ }_{-}(n-1,t+1)}{\rho(n,t)},\\
\sin  \widetilde\theta^{ }_{n,t} &=& \frac{ \widetilde\psi^{ }_{-}(n,t)  \widetilde\psi^{ }_{+}(n+1,t+1)}{\rho(n,t)}\nonumber\\
&+& \frac{ \widetilde\psi^{ }_{+}(n,t) \widetilde \psi^{ }_{-}(n-1,t+1)}{\rho(n,t)}.
\end{eqnarray*}  
The Cauchy-Schwarz inequality ensures that this quantities are well defined, i.e., $|\cos \widetilde\theta^{ }_{n,t}|\leq1$, $|\sin \widetilde\theta^{ }_{n,t}|\leq1$.



\begin{thebibliography}{00}


\bibitem{VA12} \Journal{S. E. Venegas-Andraca}{2012}{Quantum walks: a comprehensive review}{Quantum Inf. Process.}{11}{1015}{106} 

\bibitem{GW94} \Book{G. H. Weiss}{1994}{Aspects and Applications of the Random Walk}{North Holland}{New York}

\bibitem{ADZ93} \Journal{Y. Aharonov, L. Davidovich, and N. Zagury}{1993}{Quantum random walks}{Phys. Rev. {\rm A}}{48}{1687}{90} 

\bibitem{TM02} \JournalE{B. C. Travaglione and G. J. Milburn}{2002}{Implementing the quantum random walk}{Phys. Rev. {\rm A}}{65}{032310} 

\bibitem{NK03} \Journal{N. Konno}{2003}{Quantum random walks in one dimension}{Quantum Inf. Process.}{1}{345}{54} 

\bibitem{JK03} \Journal{J. Kempe}{2003}{Quantum random walks: An introductory overview}{Contemp. Phys.}{44}{307}{27} 


\bibitem{ABNVW01} \Bookwd{A. Ambainis, E. Bach, A. Nayak, A. Vishwanath, and J. Watrous}{2001}{{\rm in} One Dimensional Quantum Walks, {\rm Proceedings of the thirty-third annual ACM symposium on Theory of Computing}}{ACM New York}{New York}, p. 37. 



\bibitem{BFT08} \JournalE{D. Bulger, J. Freckleton, and J. Twamley}{2008}{Position-dependent and cooperative quantum Parrondo walks}{New J. Phys.}{10}{093014} 

\bibitem{SK10} \JournalE{Y. Shikano and H. Katsura}{2010}{Localization and fractality in inhomogeneous quantum walks with self-duality}{Phys. Rev. {\rm E}}{82}{031122} 

\bibitem{KW11} \JournalE{P. Kurzy\'nski and A. W\'ojcik}{2011}{Discrete-time quantum walk approach to state transfer}{Phys. Rev. {\rm A}}{83}{062315} 

\bibitem{CGMV12} \Journal{M. J. Cantero, F. A. Gr\"unbaum, L. Moral, and L. Vel\'azquez}{2012}{The CGMV method for quantum walks}{Quantum Inf. Process.}{11}{1149}{92} 

\bibitem{KLS13} \Journal{N. Konno, T. \L uczak, and E. Segawa}{2013}{Limit measures of inhomogeneous discrete-time quantum walks in one dimension}{Quantum Inf. Process.}{12}{33}{53} 

\bibitem{ZXT14} \JournalE{R. Zhang, P. Xue, and J. Twamley}{2014}{One-dimensional quantum walks with single-point phase defects}{Phys. Rev. {\rm A}}{89}{042317} 

\bibitem{XQTS14} \JournalE{P. Xue, H. Qin, B. Tang, and B. C. Sanders}{2014}{Observation of quasiperiodic dynamics in a one-dimensional quantum walk of single photons in space}{New J. Phys.}{16}{053009} 

\bibitem{AZ16} \Journal{A. Suzuki}{2016}{Asymptotic velocity of a position-dependent quantum walk}{Quantum Inf. Process.}{15}{103}{19} 


\bibitem{RMM04} \JournalE{P. Ribeiro, P. Milman, and R. Mosseri}{2004}{Aperiodic Quantum Random Walks}{Phys. Rev. Lett.}{93}{190503} 

\bibitem{BNPRS06} \JournalE{M. C. Ba\~nuls, C. Navarrete, A. P\'erez, E. Rold\'an, and J. C. Soriano}{2006}{Quantum walk with a time-dependent coin}{Phys. Rev. {\rm A}}{73}{062304} 

\bibitem{AR09a} \Journal{A. Romanelli}{2009}{The Fibonacci quantum walk and its classical trace map}{Physica {\rm A}}{388}{3985}{90}

\bibitem{AR09b} \JournalE{A. Romanelli}{2009}{Driving quantum-walk spreading with the coin operator}{Phys. Rev. {\rm A}}{80}{042332} 

\bibitem{MM14} \JournalE{M. Montero}{2014}{Invariance in quantum walks with time-dependent coin operators}{Phys. Rev. {\rm A}}{90}{062312} 


\bibitem{FAJ04} \Journal{A. P. Flitney, D. Abbott, and N. F. Johnson}{2004}{Quantum walks with history dependence}{J. Phys. {\rm A}}{37}{7581}{91} 

\bibitem{RBG13} \JournalE{P. P. Rohde, G. K. Brennen, and A. Gilchrist}{2013}{Quantum walks with memory provided by recycled coins and a memory of the coin-flip history}{Phys. Rev. {\rm A}}{87}{052302} 

\bibitem{SWH14} \JournalE{Y. Shikano, T. Wada, and J. Horikawa}{2014}{Discrete-time quantum walk with feed-forward quantum coin}{Sci. Rep.}{4}{4427} 


\bibitem{JM10} \Journal{A. Joye and M. Merkli}{2010}{Dynamical Localization of Quantum Walks in Random Environments}{J. Stat. Phys.}{140}{1025}{53} 

\bibitem{AJ11} \Journal{A. Joye}{2011}{Random Time-Dependent Quantum Walks}{Commun. Math. Phys.}{307}{65}{100} 

\bibitem{AVWW11} \JournalE{A. Ahlbrecht, H. Vogts, A. H. Werner, and R. F. Werner}{2011}{Asymptotic evolution of quantum walks with random coin}{J. Math. Phys.}{52}{042201} 

\bibitem{ACMSWW12} \Journal{A. Ahlbrecht, C. Cedzich, V. B. Scholz, A. H. Werner, and R. F. Werner}{2012}{Asymptotic behavior of quantum walks with spatio-temporal coin fluctuations}{Quantum Inf. Process.}{11}{1219}{49} 

\bibitem{AJ12} \Journal{A. Joye}{2012}{Dynamical localization for d-dimensional random quantum walks}{Quantum Inf. Process.}{11}{1251}{69} 


\bibitem{RM01} \Journal{R. Metzler}{2001}{Non-homogeneous random walks, generalised master equations, fractional Fokker-Planck  equations, and the generalised Kramers-Moyal expansion}{Eur. Phys. J. \text{B}}{19}{249}{58} 

\bibitem{FS07} \JournalE{O. Flomenbom and R. J. Silbey}{2007}{Path-probability density functions for semi-Markovian random walks}{Phys. Rev. {\rm E}}{76}{041101}

\bibitem{LRT13} \Journal{P. Lafitte-Godillon, K. Raschel, and V. C. Tran}{2013}{Extinction probabilities for a distylous plant population modeled by an inhomogeneous random walk on the positive quadrant}{SIAM J. Appl. Math.}{73}{700}{22} 

\bibitem{MPW17} \Book{M. Menshikov, S. Popov, and A. Wade}{2017}{Non-homogeneous Random Walks: Lyapunov Function Methods for Near-Critical Stochastic Systems}{Cambridge University Press}{New York} 


\bibitem{MM16} \JournalE{M. Montero}{2016}{Classical-like behavior in quantum walks with inhomogeneous, time-dependent coin operators}{Phys. Rev. {\rm A}}{93}{062316} 


\bibitem{MM15} \Journal{M. Montero}{2015}{Quantum walk with a general coin: exact solution and asymptotic properties}{Quantum Inf. Process.}{14}{839}{66} 

\bibitem{NK05} \Journal{N. Konno}{2005}{A new type of limit theorems for the one-dimensional quantum random walk}{J. Math. Soc. Japan}{57}{1179}{95} 

\bibitem{BP07} A. Bressler and R. Pemantle in \textit{Quantum random walks in one dimension via generating functions}, Proceedings of the 2007 Conference on Analysis of Algorithms, (2007),  pp. 403. 

\bibitem{BSJ15} \JournalE{I. Bezd\v{e}kov\'a, M. \v{S}tefa\v{n}\'ak, and I. Jex}{2015}{Suitable bases for quantum walks with Wigner coins}{Phys. Rev. {\rm A}}{92}{022347} 

\bibitem{SJ16} \JournalE{M. \v{S}tefa\v{n}\'ak, and I. Jex}{2016}{Persistence of unvisited sites in quantum walks on a line}{Phys. Rev. {\rm A}}{93}{032321} 




\end{thebibliography}
\end{document}